\documentstyle[prl,aps,epsf]{revtex}
\begin{document}
\twocolumn[\hsize\textwidth\columnwidth\hsize\csname@twocolumnfalse\endcsname
\title{
Orbital ferromagnetism and anomalous Hall effect in antiferromagnets  
on distorted  fcc lattice
}
\author{Ryuichi Shindou$^1$ and Naoto Nagaosa$^{1,2}$}
\address{$^1$ Department of Applied Physics, University of Tokyo,
Bunkyo-ku, Tokyo 113-8656, Japan}
\address{$^2$ Correlated Electron Research Center, Tsukuba, Ibaraki 
303-0046, Japan}
\date{May,1,2001}
\maketitle
\begin{abstract}
The Berry phase due to the spin wavefunction gives rise to the 
orbital ferromagnetism and anomalous Hall effect in the non-coplanar 
antiferromagnetic ordered state on face-centered-cubic (fcc) 
lattice once the crystal is distorted perpendicular to  (1,1,1)
or (1,1,0)- plane.  
The relevance to the real systems $\gamma$-FeMn and NiS$_2$ 
is also discussed.
\end{abstract}
\
]
It has been recognized for a long term that the 
chirality plays  important roles in the physics of  
frustrated spin systems 
\cite{villain,miyashita,kawamura,laughlin,wen,nagaosa}. 
These degrees of freedom are distinct from 
the (staggered) magnetization, and could show phase transition 
without  magnetic ordering\cite{villain,miyashita,kawamura}.
Especially since the discovery of the high-Tc cuprates, 
the scalar spin chirality 
\begin{eqnarray}
\chi_{ijk} =  {\vec S}_i \cdot 
( {\vec S}_j \times {\vec S}_k )\label{1-01}  
\end{eqnarray}
has been a key theoretical concept in the physics of 
strongly correlated electronic systems \cite{laughlin,wen,nagaosa}.
This spin chirality acts as the gauge flux for the 
charge carriers moving in the background of the fluctuating spins.
In order for the spin chirality $\chi_{ijk}$ to be ordered,
both the time-reversal (T) and parity (P) symmetries
must be broken.  Broken T and P symmetries in 2D bring about many
intriguing physics such as parity anomaly \cite{jak,ninomiya}
, anyon superconductivity \cite{anyon},  and
quantized Hall effect {\it without} external magnetic field 
\cite{haldane}.  A physical 
realization of the last one  has been discussed \cite{ohgushi}
in the context of anomalous Hall effect (AHE) in 
ferromagnets via the  spin chirality mechanism 
\cite{ong,salamon,ye,taguchi}.

In this paper we explore the chiral spin state in the
ordered antiferromagnet (AF) on the three-dimensional 
face-centered-cubic (fcc) lattice.
The AF on the fcc lattice is a typical frustrated system,
and nontrivial spin structure with the finite spin chirality 
in eq.(1) is expected. For example, in  the charge transfer (CT) insulator
NiS$_2$ \cite{nis} and in the metallic alloy 
$\gamma$-FeMn \cite{endoh} the non-coplanar
spin structure (so-called triple-Q structure shown in Fig.\ 1a) has 
been observed. A theoretical explanation for this 
structure is the following.
Let us consider the case where the 
lattice points are divided into 4-sublattices as shown in Fig.\ 1a.
Denoting the (classical) spin moment at each sublattice as 
${\vec S}_a$ ($a = 1,2,3,4$), the 2-spin exchange interaction energy 
is written as $H_2 \propto ( \sum_{a=1,4} {\vec S}_a )^2$.
Therefore the condition of the lowest energy 
$\sum_{a=1,4} {\vec S}_a ={ \vec 0}$
does not determine the 
spin structure and leaves many degenerate lowest energy 
configurations.
Then the interactions which lift this degeneracy 
such as the 4-spin exchange interaction 
become important\cite{yoshida,yoshimori}.
In paticular the phenomenological Ginzburg-Landau theory 
for the 4-spin exchange interaction is given as 
$H_4 = J_4 \sum_{a \ne b} ( {\vec S}_a \cdot {\vec S}_b )^2$
\cite{yoshimori}.
With positive $J_4$, the ground state configuration
is given by 
${\vec S}_1 = ( {1 \over {\sqrt{3}} }, { 1 \over {\sqrt{3} }},
{ 1 \over {\sqrt{3}}} )$,
${\vec S}_2 = ( {1 \over {\sqrt{3}}}, -{ 1 \over {\sqrt{3}}},
-{ 1 \over {\sqrt{3}}} )$,
${\vec S}_3 = ( -{1 \over {\sqrt{3}}}, { 1 \over {\sqrt{3}}},
-{ 1 \over {\sqrt{3}}} )$,
${\vec S}_4 = ( -{1 \over {\sqrt{3}}}, -{ 1 \over {\sqrt{3}}},
{ 1 \over {\sqrt{3}}} )$, where 
each direction corresponds to the four corners from the 
center of a tetrahedron. 
This non-coplanar spin structure gives the scalar chirality in eq.(1)
locally.
For the itinerant system $\gamma$-FeMn, a recent band structure 
calculation \cite{sakuma}
concluded the stability of the triple-Q spin strucure, in 
agreement with experiment \cite{endoh}.
\input epsf
\begin{figure}[h]
\begin{center}
\epsfysize=5.5cm
\epsfbox{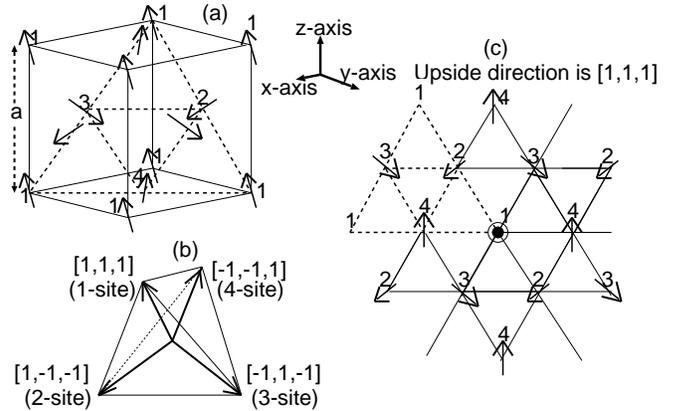}
\end{center}
\caption{(a) Triple-Q spin structure on fcc lattice. (b)
Relation of the 4-spin moments ${\vec S}_a$ ($a=1,2,3,4$).
(c) Triangle lattice as a cross section of he fcc lattice
perpendicular to [1,1,1]-direction.
}
\label{1}
\end{figure}

Experimentally  anomalous behaviors in the fcc 
AF are often observed.
For example there occurs  mysterious weak 
ferromagnetism (WF) in NiS$_2$ below the second AF 
transition temperature $T_{N2}$
\cite{ferro}.
The Hall effect in this material is also large and strongly 
temperature dependent \cite{thio}.
In Co(S$_x$Se$_{1-x}$)$_2$, the AHE is 
enhanced in the intermediate $x$ region, where the nontrivial 
magnetism is realized \cite{adachi}, and CeSb shows the largest
Faraday rotation ever observed \cite{cesb}.

As shown below, the triple-Q AF state (Fig.\ 1a) on 
the distorted fcc lattice gives the orbital ferromagnetism
accompanied by the AHE 
and even the realization of the three-dimensional 
quantum Hall liquid \cite{halperin} {\it without} the external 
magnetic field.

The Hubbard on-site repulsive interaction $U$ can be decomposed in 
terms of the Stratonovich-Hubbard transformation as
\begin{eqnarray}
{  U \over 2} {\vec \varphi}_i^2 -  U {\vec \varphi}_i \cdot 
c^\dagger_{i \alpha} {\vec \sigma}_{\alpha \beta} c_{i \beta},
\label{1-02} 
\end{eqnarray}
where $\vec{\sigma}$ are the Pauli matrices and ${\vec \varphi_i}$ 
is the spin fluctuation field. 
This field acts as the local effective magnetic field
for the electron , which can be regarded as the c-number
in the mean field approximation for the magnetically ordered state. 
The symmetry properties of the ground state is captured by this
approximation , and the quantized Hall effect discussed below is  
stable against the fluctuation of the ${\vec \varphi}$-field as long 
as the gap does not collapse. Therefore we can consider this system 
as the free electron model in the background of the static field  
$U {\vec \varphi}_i$. In paticular when this field 
$U |{\vec \varphi}_i|$ is 
much stronger than the transfer integrals $t_{ij}$'s, we can 
map the model into that of spinless fermions 
$H = - \sum_{ij} t_{ij}^{\rm eff} f^\dagger_i f_j $
with the effective transfer integral 
\begin{eqnarray}
t_{ij}^{\rm eff} &=& t_{ij} {\langle}\chi_i | \chi_j {\rangle}\nonumber \\
&=&t_{ij} (\cos { {\theta_i} \over 2} \cos { {\theta_j} \over 2}
+\sin { {\theta_i} \over 2} \sin { {\theta_i} \over 2}
e^{ - i ( \phi_i - \phi_j )}),\label{1-03}
\end{eqnarray}
where $| \chi_i {\rangle}=  [ \cos { {\theta_i} \over 2} ,
\sin { {\theta_i} \over 2}e^{ i \phi_i} ]^t $
is the spin wavefunction and $\theta_i$ and $\phi_i$ are the
polar coordinates of the spin direction.
The phase factor appearing in $t_{ij}^{\rm eff}$
can be regarded as the gauge vector potential $a_{\mu}(\vec{r})$
, and the corresponding gauge flux is related to $\chi_{ijk}$ in 
eq.(\ref{1-01})\cite{wen,nagaosa}.
 However, it should be noted here that the 
effective magnetic flux vanishes  when averaged over the unit cell 
because of the periodicity of the vector potential
$a_\mu(\vec r) = a_\mu({\vec r} + {\vec T}_i)$
(${\vec T}_i:$ primitive vector).
Therefore, the net effect of the gauge flux comes from the 
fact that there are more than two atoms and/or orbitals 
in the unit cell and the resultant multiband structure.
Then each band is characterized by the 
Chern number \cite{tknn}. The Chern number
appears as a result of the spin-orbit interaction and/or the
spin chirality in ferromagnets \cite{ohgushi,onoda}. 
In ferromagnets the T-broken symmetry is manifest, while
in AF the T-operation combined with the 
translation operation often constitutes the unbroken symmetry. 
In the latter case, the nonzero Hall conductivity $\sigma_{xy}$ 
is forbidden. However when there are more than two sublattices and 
the spin structure is non-collinear, this combined symmetry
would be absent and finite $\sigma_{xy} $ is not forbidden.

In order to facilitate the understanding, first consider 
the triangular lattice, which is the
(1,1,1)-cross section of the fcc lattice (Fig.\ 1c), 
and tight binding model with nearest neighbor hopping given
in eq.(\ref{1-03}) has the gauge flux of $\pi/2$ penetrating each triangle.
The unit cell is composed of eight triangles including four atoms.
Thus the total gauge flux 
penetrating the unit cell is $4 \pi \equiv 0 ({\rm{Mod}}(2 \pi))$, 
which is consisitent with
the periodicity of the lattice. 
The Chern number can be estimated  analytically
 as  $-\frac{e^2}{h}$($\frac{e^2}{h}$) 
for the lower (higher) energy band.
Therefore this two-dimensional spin configuration on 
the triangular lattice offers an example where
the spin chirality orders without the ferromagnetic spin moment.

When one considers the three-dimensional fcc lattice,
on the other hand, there are three other cross sections equivalent to
[1,1,1]-direction, namely [-1,-1,1],[-1,1,-1],[1,-1,-1].
Therefore it is naturally expected that 
the net spin chirality is zero, because the spin chiralities
are the vector quantities and the sum of these four vectors is 
zero. Actually $\sigma_{xy} = \sigma_{yz} = \sigma_{zx} = 0$ 
for the fcc lattice as shown below.
However, this means that the chirality remains finite when 
the symmetry between the four directions is violated.
For example, when the lattice is distorted along 
[1,1,1]-direction, it is expected 
that $\sigma_H = \sigma_{xy} = \sigma_{yz}
= \sigma_{zx} $ becomes finite.
We express the distortion along [1,1,1]-direction by putting 
the transfer integral within the (1,1,1)-plane as
$t_{\rm intra} = 1$, while that between the planes as 
$t_{\rm inter} = 1-d$ \cite{comment}. 
As the unit cell is cubic shown in 
Fig.\ 1a , the first Brillouin zone (BZ) is cubic
: $[-\frac{\pi}{a},\frac{\pi}{a}]^{3}$. From now on, we set $a=1$ .
Then the Hamiltonian matrix $H(\vec{k})$ for each $\vec k$ 
is given by 
\begin{eqnarray}
H(\vec{k}) =
     \left(
      \begin{array}{cccc}
      0 & e^{-i\frac{\pi}{6}}f_{2} & e^{i\frac{\pi}{6}}f_{1}  
      & f_{3} \\
      e^{i\frac{\pi}{6}}f_{2}  & 0 & e^{-i\frac{\pi}{6}}f_{3}
      & e^{i\frac{2\pi}{3}}f_{1} \\
      e^{-i\frac{\pi}{6}}f_{1} & e^{i\frac{\pi}{6}}f_{3} & 0
      &  e^{-i\frac{2\pi}{3}}f_{2}\\
      f_{3} & e^{-i\frac{2\pi}{3}}f_{1} 
      & e^{i\frac{2\pi}{3}}f_{2} & 0
      \end{array}
      \right),  \label{1-1}
\end{eqnarray}
where $f_{1}=2(1-d)\cos(\frac{k_{z}}{2}+\frac{k_{x}}{2})
+2\cos(-\frac{k_{z}}{2}+\frac{k_{x}}{2})$, $f_{2}
=2(1-d)\cos(\frac{k_{x}}{2}+\frac{k_{y}}{2})
+2\cos(-\frac{k_{x}}{2}+\frac{k_{y}}{2})$, $f_{3}
=2(1-d)\cos(\frac{k_{y}}{2}+\frac{k_{z}}{2})
+2\cos(-\frac{k_{y}}{2}+\frac{k_{z}}{2})$.
In this Hamiltonian, two bands 
($\varepsilon_{1,2} = -\sqrt{f_{1}+f_{2}+f_{3}}$) and 
upper two 
bands ($ \varepsilon_{3,4}= \sqrt{f_{1}+f_{2}+f_{3}}$) are 
completely degenerate. 
At $d$ = 0, these two dispersions touch along the edge of the 
1st BZ, i.e. 
$(k_x = \pm\pi, k_y = \pm \pi, k_z)$, $(k_x , k_y 
= \pm \pi, k_z=\pm \pi)$, $(k_x 
= \pm\pi, k_y , k_z=\pm \pi)$. Fixing $k_z$, for 
example, the $\vec k$-point $(k_x = \pm \pi,k_y = \pm \pi)$ 
is the center of the massless Dirac fermion
(Weyl fermion) in (2+1)D. Therefore the band touching along the 
edge can be regarded as an assemble of the (2+1)D Weyl fermions.
What happens for finite $d$ differs for positive and negative
values of $d$. For $d>0$ (elongation along [1,1,1]-direction),
all the Weyl fermions along the edges open a gap and turn into
the massive Dirac fermions. Therefore the gap opens in the density 
of states  centered at $\varepsilon = 0$. 
For $d<0$ (suppression along [1,1,1]-direction), on the other hand,
all the (2+1)D Weyl fermions along the edges open 
the gap; two (3+1)D Weyl fermions emerge instead at 
$(k_{x},k_{y},k_{z})={\pm}2\arcsin(\sqrt{\frac{4-2d}{4-4d}})(1,1,1)
={\pm}D(1,1,1)$
where $\pm$ correspond to right- and left-handed chirality 
\cite{ninomiya}.
Here we consider the mass of the Dirac fermions at the edges of the 
BZ. For $d>0$, the mass of the (2+1)D fermion 
along the $k_\mu$-axis is positive for every $k_\mu$, while 
that changes sign at $k_\mu = \pm D$ for $d<0$.

Based on this observation, the transverse conductivity can be
calculated in terms of the formula in \cite{tknn} 
\begin{eqnarray}
\sigma_{xy}
&=&\sum_{n=1}^4 \int_{[-\pi:\pi]}\frac{dk_{z}}{2\pi}
\int_{[-\pi:\pi]^{2}}
\frac{d k_x d k_y}{2{\pi}i} f(\varepsilon_n (\vec k) ) 
(\nabla_{\vec{k}} \times
\vec{A}_n(\vec{k}))_{z}\nonumber \\
&=&\frac{e^{2}}{h}\int_{[-\pi:\pi]}\frac{dk_{z}}{2\pi} 
\sigma_{xy}(k_{z}),\label{1-11}
\end{eqnarray}
where $f(\varepsilon)$ is the Fermi distribution function, and
$\vec{A}_n (\vec{k})={\langle} n \vec{k} |{\nabla}_{\vec{k}}
| n \vec{k} {\rangle}$  is the vector potential created by 
the Bloch wavefunction $| n {\vec k}{\rangle}$.
Let us consider $\mu = 0$ case, i.e. quarter filled case. 
$\sigma_{xy}(k_z)$ is the 
Hall conductance in (2+1)D system with fixed $k_z$, which is 
determined by the parity anomaly caused by the Dirac fermions 
along the edge of the BZ. Explicit caluculation shows at $T = 0\rm{K}$ ,
$\sigma_{xy}(k_z)=-\frac{e^{2}}{h}{\rm{sign}}(f_{2}(k_z)\cdot{d})$
where $
f_{2}(k_z){\equiv}(4-2d)\cos(\frac{k_z}{2})^{2}+2d{\cdot}
\sin(\frac{k_z}{2})^{2}$.
For $d>0$, $\sigma_{xy}(k_z) =- { {e^2} \over {h} }$ for all $k_z$, 
while $\sigma_{xy}(k_z) = \frac{e^2}{h}$ for 
$k_{z}=(-D,D)$ and $\sigma_{xy}(k_z) = -\frac{e^2}{h}$  
for $k_{z}=[-\pi,-D) $ and $(D,\pi]$ in the case of nagative $d<0$.
Integrating over $k_z$, we obtain 
$\sigma_{xy} =- { {e^2} \over h }$ for $d>0$ and
$\sigma_{xy} =\frac{e^2}{h}\biggl[\frac{4}{\pi}
\arcsin(\sqrt{\frac{4-2d}{4-4d}})-1\biggr]$ for $d<0$, while
$\sigma_{xy}$ vanish for $d = 0$. One can easily check that
$\sigma_{xy} = \sigma_{yz} = \sigma_{zx} = \sigma_H$, and that the 
sign of $\sigma_H$ changes when all the spins ${\vec S}_a$ are
inverted ($ {\vec S}_a \to - {\vec S}_a$).
One can confirm  analytically in eq.(\ref{1-1}) that 
the cancellation between the contributions from the 
two $\vec k$-points 
related by the parity (P) symmetry, e.g. $(k_{x},k_{y},k_{z})$ 
and, $(-k_{x},k_{y},k_{z})$, occurs and $\sigma_{xy}$ 
is zero for the undistorted fcc lattice. This remains true 
even when we modify ${\vec S}_a$ from those given 
in Fig.\ 1b. For example even with finite 
magnetization  ($\sum_{a = 1}^4 {\vec S}_a
\ne {\vec 0}$), $\sigma_{xy}(\omega) = 0$ for any $\omega$ on the
undistorted fcc lattice. In short, P-symmetry is not broken while
T-symmetry is broken in the undirtorted fcc lattice, where 
finite $\sigma_H$ is forbidden.
An distortion along [1,1,1]-direction {\it does} break this 
P-symmetry and produces finite $\sigma_H$.
%

\begin{figure}
\begin{center}
\epsfysize=5.5cm\epsfbox{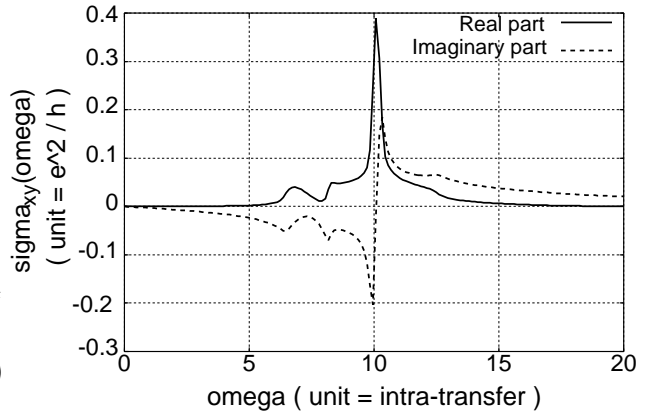}
\end{center}
\caption{$\sigma_{H}$($\omega$) for $d = - 0.1 , U 
\vert{\vec{\varphi}_{i}}
\vert = 5 $ }
\label{6}
\end{figure}

Next we turn to the half filled case.
\noindent When the chemical potential is in the Mott gap, 
i.e. Mott insulator , the d.c. $\sigma_H$ vanishes. 
However the optical $\sigma_H(\omega)$ can be finite,
which corresponds to the Faraday and/or the Kerr rotation.
In order to calculate $\sigma_H(\omega)$, we have to 
take into account optical transitions between the lower and the 
upper Hubbard bands. In the mean field picture, this corresponds
to the two splitted bands due to the effective magnetic field
$U \vec \varphi_i$ in eq.(2).
Diagonalizing the tight-binding Hamiltonian with this local 
``magnetic field'', we can calculate  $\sigma_H(\omega)$.
$\sigma_H(\omega)$ shown in Fig.\ {\ref{6}} is for $d=-0.1,U 
\vert{\vec{\varphi}_{i}}\vert = 5 $.

Now we turn to the orbital ferromagnetism induced by the 
distortion. Orbital magnetization $M$ is given by  
\begin{eqnarray}
M &=& \lim_{B {\rightarrow} 0}
\int_{-\infty}^{\epsilon_{F}}\frac{{\partial}N(B,\mu)}{{\partial}B}
d\mu.\label{1-05} 
\end{eqnarray}
In terms of the St${\rm{\check{r}}}$eda formula
${\cite{streda}}$, i.e. $\sigma_{xy}|_{\mu}=\sigma^{I}_{xy}|_\mu +ec
\frac{{\partial}N(B,\mu)}{{\partial}B}$ with
$\sigma_{xy}^{I} |_\mu=i\frac{1}{2h}{\rm{Tr}}[\hat{J}_{x}G^{+}
(\mu)\hat{J}_{y}\delta(\mu -  H) -\hat{J}_{x}
\delta(\mu - H)\hat{J}_{y}G^{-}(\mu)]$, we estimate the integrand 
in eq.(\ref{1-05}) from ${\sigma}_{xy}|_{\mu}$ and
 ${\sigma}^{I}_{xy}|_{\mu}$. 
\begin{figure}
\begin{center}
\epsfysize=5.5cm\epsfbox{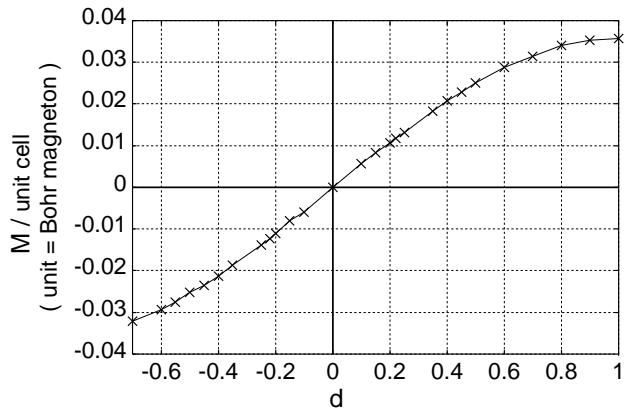}
\end{center}
\caption{Magnetization $M$ as a function of $d$, where  $U 
\vert{\vec{\varphi}_{i}}\vert = 5 $  
( we set $\frac{h^{2}}{ma^{2}} {\sim}t_{\rm intra}=1$ ) }
\label{7}
\end{figure}
\noindent Shown in Fig.\ 3 is the orbital magnetization $M$ as a function of
the distortion $d$ for the half filled case. The sign of orbital
 magnetization $M$ changes when all the spins 
are inverted. (Magnetization depicted in Fig.\ 3 corresponds to the 
triple-Q structure in Fig. 1a,b.)
Note that $M$ is finite even though  the d.c. $\sigma_H=0$ in this case,
because the former is determined by the integral over the 
occupied states.

 In real materials, the configuration with all the spins being inverted 
 (which means that the chirality is also inverted) 
has the same energy. Thus it is expected that the domain structure of these 
two chiralities is formed. In order to align this chiral domain, we 
must cool down into AF order phase by applying external strain 
{\it and} external magnetic field, which couples to the orbital
 magnetization and prefers one of the chiral domain. (For $d>0$, the
magnetic field in [1,1,1]-direction prefers the triple-Q structure as
in Fig.\ 1a,b.) Lastly it is noted here that 
the distortion along [1,1,0] or other three equivalent 
directions produces the similar effect as
discussed above, while that along [1,0,0] etc. does not.
This also offers a way to test this idea experimentally. 

Now we discuss the experimental relevance of these results.
One of the most promising materials for this spin chirality
mechanism is the itinerant AF 
$\gamma$-Fe$_x$Mn$_{1-x}$ alloy as mentioned above \cite{endoh}. 
In this material, the triple-Q structure is observed for $0.35<x<0.8$.
 This material remains metallic even below $T_{N}$.
Because the crystal structure is the undistorted fcc, one needs
to apply uniaxial stress  toward the [1,1,1] or [1,1,0]-direction. 
Although the band structure is rather complicated and there is no
gap in the density of states \cite{sakuma}, we expect the 
transverse conductivity $\sigma_H$ of the order of 
$e^2 d/ ha \cong 1400 d \Omega^{-1} {\rm{cm}}^{-1} $ 
for $a = 3.6 \rm{\AA} $ when the 
chiral domains are aligned by the field cooling.

Another candidate is the CT insulator AF NiS$_2$.
In this material the valence of Ni is 2+, and
the angular orbital moment is quenched with $d^8$ configuration.
Thus spin-orbit interaction is expected to be 
small, and this material is an ideal laboratory 
to study the spin chirality mechanism. 
It shows two successive magnetic phase transitions at 
$T_{N1} = 40\rm{K}$ and $T_{N2} = 30\rm{K}$.
Between $T_{N1}$ and $T_{N2}$, the magnetic strucure is 
given by Fig.\ 1a,b (type I AF state). 
At $T_{N2}$ the type II AF structure appears in addition 
to the type I structure. This is accompanied 
by the rhombohedral distortion and with the mysterious WF \cite{ferro}.
This lattice distortion ($d>0$) lifts 
the quasi-degeneracy of type I and
type II structures \cite{matsuura}. This brings about 
finite orbital ferromagnetism, 
which can be detected by hysteresis in the magnetization 
curve under the magnetic field. 
Therefore the present results give a possible 
scenario for the WF in NiS$_2$.  
Even in the temperature range between  $T_{N1}$ and $T_{N2}$, 
where lattice is not distorted spontaneously
, suppression toward [1,1,1] or [1,1,0]-direction 
is predicted to bring about an orbital 
ferromagnetism. (The compressibility of this material is of 
the order of $10^{-3}{\rm{kbar}}^{-1}$\cite{miya} and we 
need $\sim$\noindent 140kbar   uniaxial pressure  to 
produce $d=-0.1$ for example.)  Although the d.c. $\sigma_H(\omega)$  
is zero at $T = 0\rm{K}$, the optical  
$\sigma_H(\omega)$ is expected to be finite, as shown in 
Fig.\ 2. As for other fcc AF such 
as Co(S$_x$Se$_{1-x}$)$_2$, detailed studies on the 
spin structure by neutron scattering are  highly desirable, 
with which the mechanism of the Hall effect can 
be dictated.

The authors acknowledges Y. Tokura, Y. Endoh, A. Mishchenko,
M. Onoda and Y.Taguchi for fruitful discussions. 
N.N.  is supported by 
Priority Areas Grants 
and Grant-in-Aid for COE research 
from the Ministry of Education, Culture, Sports, Science and   
Technology of Japan.


\begin{references}
\bibitem{villain}J. Villain,
J. Phys. C {\bf 10}, 1717 and 4793 (1977).
\bibitem{miyashita}S.Miyashita and H.Shiba,
J. Phys. Soc. Jpn. {\bf 53}, 1145 (1984); 
Prog. Theor. Phys. Suppl. {\bf{87}} 112 (1986).
\bibitem{kawamura}H. Kawamura, Phys. Rev. Lett. {\bf 80}, 5421 (1998).
\bibitem{laughlin}V. Kalmeyer and R. B. Laughlin, Phys. Rev. Lett.
{\bf 59}, 2095 (1987); R. B. Laughlin, {\it ibid} {\bf 60}, 2677 (1998). 
\bibitem{wen}X.~G.~Wen, F.~Wilczek and A.~Zee,
Phys. Rev. {\bf B39}, 11413 (1989).
\bibitem{nagaosa}P.~A.~Lee and N.~Nagaosa,  Phys. Rev.
{\bf B46}, 5621 (1992).
\bibitem{jak} 
S.~Deser, R.~Jakiew, and S.~Templeton, 
Phys. Rev. Lett. {\bf 48}, 975 (1982); 
R.~Jakiew, Phys. Rev. D {\bf 29}, 2375 (1984).
\bibitem{ninomiya} 
H.~B.~Nielson and M.~Ninomiya, Phys. Lett. {\bf 130B}, 
389 (1983);  
\bibitem{anyon}R. B. Laughlin, 
Phys. Rev. Lett.{\bf 60}, 2677 (1988).
\bibitem{haldane}F.~D.~M.~Haldane
Phys. Rev. Lett. {\bf 61}, 2015(1988).
\bibitem{ohgushi}K. Ohgushi, S. Murakami, and N. Nagaosa, 
Phys. Rev. B{\bf 62}, R6065 (2000).
\bibitem{ong}P. Matl et al., Phys. Rev. B{\bf 57}, 10248 (1998).
\bibitem{salamon}S. H. Chun et al.,  Phys. Rev. Lett.
{\bf 84}, 757 (2000).
\bibitem{ye}Jinwu~Ye {\it et al.}, Phys. Rev. Lett. {\bf 83}, 
3737 (1999).
\bibitem{taguchi}Y. Taguchi et al., Science {\bf 291}, 2573 (2001). 
\bibitem{nis} J. A. Wilson ,and G. D. Pitt, Phil. Mag. {\bf 23},1297 (1971);
K. Kikuchi et al., J. Phys. Soc. Jpn. {\bf 45}, 444 (1978).     
\bibitem{endoh}Y. Endoh and Y. Ishikawa, 
J. Phys. Soc. Jpn. {\bf 30}, 1614 (1971).
\bibitem{yoshida} K. Yoshida , and  S. Inagaki 
J. Phys. Soc. Jpn. {\bf 50}, 3268 (1980).
\bibitem{yoshimori} A. Yoshimori, and  S. Inagaki 
J. Phys. Soc. Jpn. {\bf 50}, 769 (1981).
\bibitem{sakuma}A. Sakuma , J. Phys. Soc. Jpn.
{\bf 69}, 3072 (2000).
\bibitem{ferro}T. Thio, J. W.Benett, and T. R. Thurston, 
Phys. Rev. B{\bf 52}, 3555 (1995).
\bibitem{thio}T. Thio and J. W. Bennett, 
Phys. Rev. {\bf B50}, 10574 (1994).
\bibitem{adachi}K. Adachi, M. Matsui, and Y. Omata, 
J. Phys. Soc. Jpn. {\bf 50}, 83 (1981).
\bibitem{cesb} R. Pittini et al., Phys. Rev. Lett. {\bf 77}, 944 (1996). 
\bibitem{halperin}
M.~Kohmoto, B.~Halperin and Y-S.~Wu, Phys. Rev. {\bf B45}, 13488
(1992);B. I. Halperin, J. Journal of Applied
Phys. 26 ( Suppl.), 1913 (1987)).
\bibitem{tknn}D.~J.~Thouless et al., Phys. Rev. Lett. {\bf 49}, 405 (1982);
M.~Kohmoto, Ann. Phys. (N.Y.) {\bf 160}, 343 (1985).
\bibitem{onoda}M. Onoda, and N. Nagaosa, to be published.
\bibitem{comment}
The change of $J_4$ due to the distortion will change ${\vec S}_i$
($i = 1 - 4$), and hence the effective transfer integral $t^{eff}_{ij}$.
However this does not give the finite $\sigma_{xy}$ as long as the bare
transfer integral $t_{ij}$ remians unchanged.
\bibitem{streda} P. St${\rm{\check{r}}}$eda , J. Phys. C. Solid State
 Phys. {\bf 15}, L717 (1982).
\bibitem{matsuura}M. Matsuura et al., unpublished.  
\bibitem{miya} S. Endo, T. Mitsui, and  T. Miyadai , Phys. Lett. 
{\bf 46A}, 29 (1973).
\
\end{references}
\end{document}